\documentclass[twoside]{ilcws08}
\usepackage[latin1]{inputenc}     
\usepackage[dvips]{graphicx,epsfig,color}       
\usepackage{wrapfig,rotating}        
\usepackage{amssymb,amsmath,array}

\begin{document}
\title{Frequency of Positron Helicity Reversal }
\author{Sabine Riemann$^1$, Andreas Sch\"alicke$^1$  and A. Ushakov$^1$
\vspace{.3cm}\\
1- DESY in Zeuthen\\
Platanenallee 6, D-15738 Zeuthen, Germany
}

\maketitle

\begin{abstract}
The ILC baseline design for the positron source is based on a helical undulator and will deliver a positron beam with a polarization of 30\% or more. In this contribution the need for fast reversal of the positron helicity is discussed.
\end{abstract}

\section{The ILC positron source}
\begin{figure}[h]
\begin{tabular}{lc}
\includegraphics[width=0.47\columnwidth]{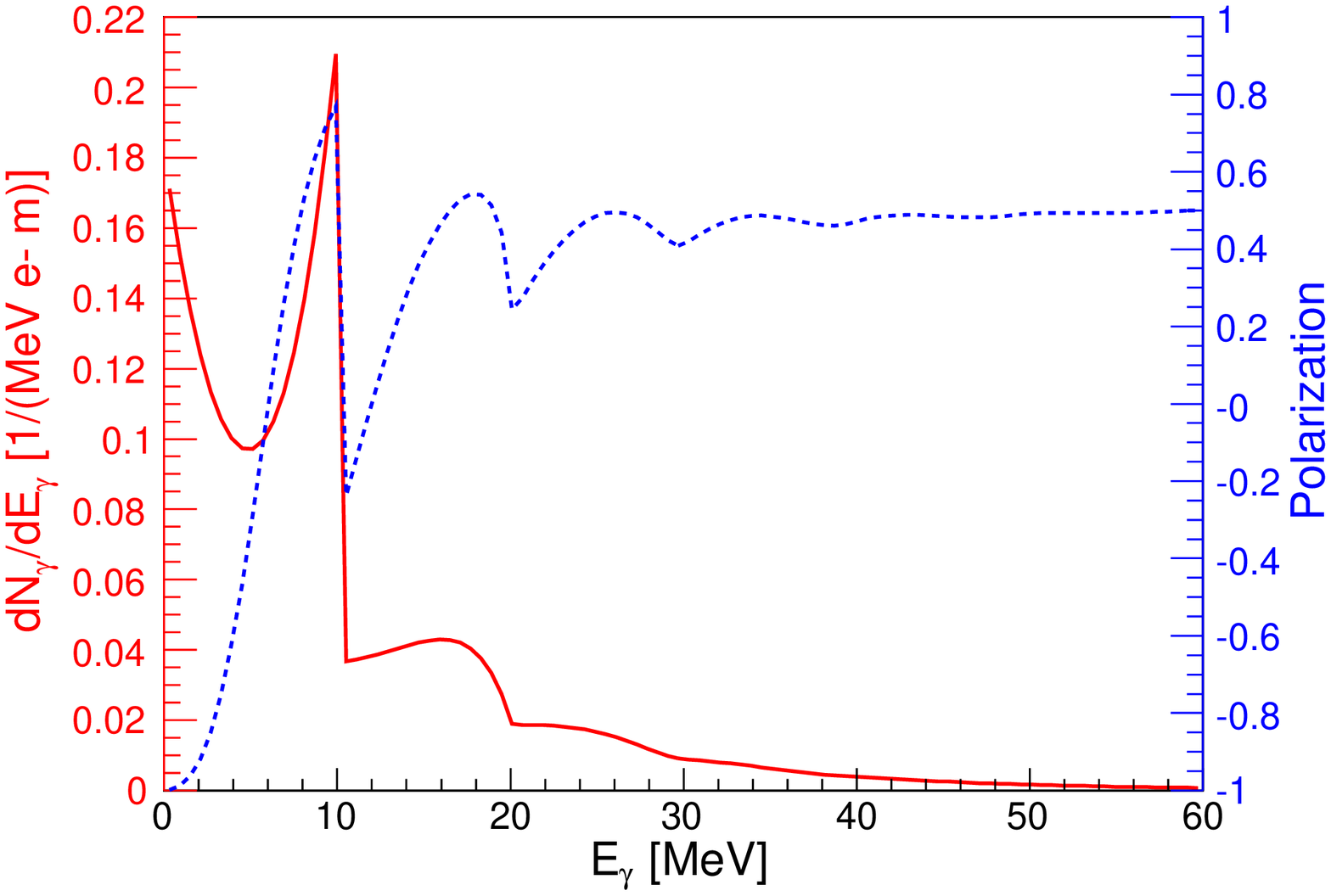}&
\includegraphics[width=0.47\columnwidth]{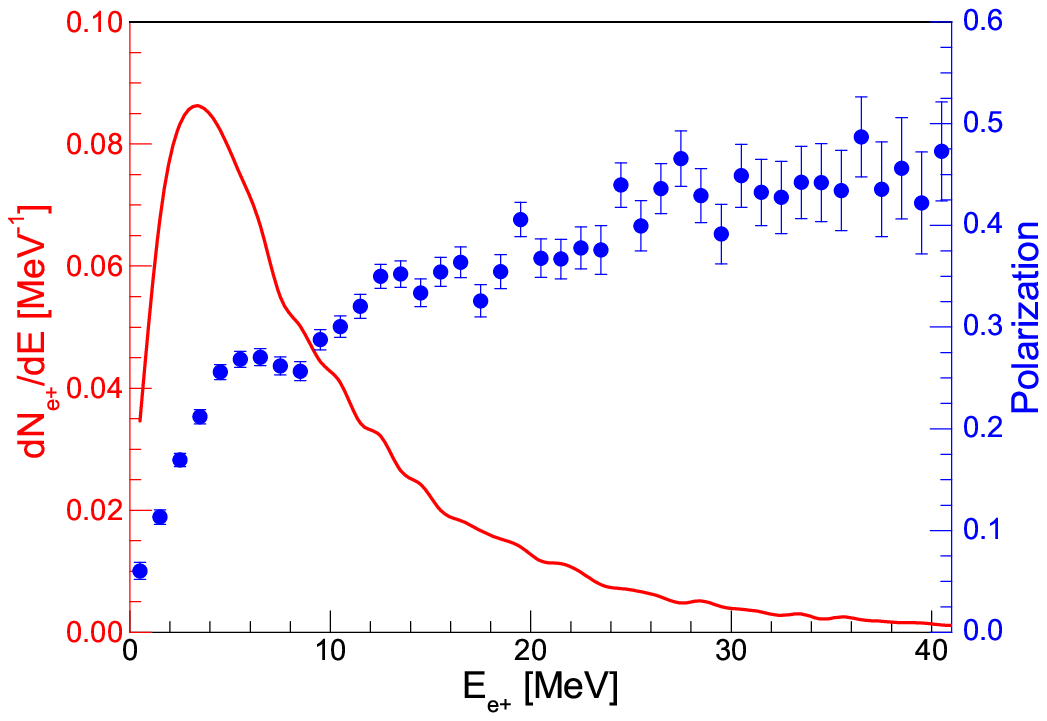}
\end{tabular}
\caption{Spectra and polarization of the photon beam (left) and the positrons after the target (right) of the ILC positron source.}\label{fig:photon}
\end{figure}

In compliance with the RDR~\cite{bib:RDRacc}, the baseline configuration 
for the positron source is based on a helical undulator passed by an electron beam of 150\,GeV. About 500\,m downstream the resulting photon beam hits a thin target, creates electron-positron pairs; the emerging positrons are collected and accelerated.
The photon beam is circularly polarized, the spectrum is shown in Fig.~\ref{fig:photon} (left plot). The circularly polarized photons create longitudinal polarized positrons whose spectrum is also presented in Fig.~\ref{fig:photon} (right plot). Taking into account the damping ring acceptance, the average positron polarization of about 30\% after the target will increase up to almost 40\%~\cite{ref:andriy}. Further increase of polarization can be achieved by collimation of the photon beam which cuts lower energetic photons with lower polarization but then the intensity loss has to be compensated, e.g. with a longer undulator.
In order to preserve the polarization, the spin vectors of electrons and positrons have to be aligned  parallel to the rotation axis of the damping ring.
Hence, in the RDR spin rotation systems \cite{ref:spin-rot} for the electron and positron line  are foreseen before and after the DR to rotate the 
spin vector from the longitudinal to the vertical direction and back.

The helicity of the electrons will be chosen randomly on a train-by-train basis. But the helicity of the positrons is defined by the undulator design, i.e.\ the orientation of the helix. A fast change of the spin vector orientation is only possible using additional spin flip equipment; design suggestions can be found in references~\cite{ref:flip5000} and \cite{ref:flip400}. Fast spin flip has not been taken into account in the RDR but it is essential for physics measurements. In this paper the need for frequent spin flip  will be illustrated.

\section{Precision measurements at the ILC }
It is a decisive advantage of an e$^+$e$^-$ collider that the initial state for the particle collision is well known; particle type, energy, and polarization are  given. Providing high luminosities, precision measurements can be performed. If both, electron and positron beam  are polarized, the helicities for the initial state can be chosen and specific studies of  physics processes of the Standard Model and beyond are possible. The role of polarization has been discussed extensively~\cite{ref:power} and having in mind the chirality of weak interactions it is clear that polarization of both beams will increase the performance of precision measurements and advance the understanding of the results.

\subsection{ILC running strategy}
The ILC described in the RDR will allow physics measurements between 200\,GeV and 500\,GeV. During the first four years of running 
an integrated luminosity of 500\,fb$^{-1}$ will be delivered. The polarization of the electron beam is at least 80\%.
The corresponding event numbers expected for important benchmark processes at the ILC during the first four years are $10^4$ for e$^+$e$^- \rightarrow~$HZ ($m_H=120\,$GeV, $\sqrt{s}=350\,$GeV),  about $10^5$ for e$^+$e$^- \rightarrow t \bar{t}$ ($\sqrt{s}\approx 350\,$GeV),  roughly $5 \times 10^5$ ($10^5$) for e$^+$e$^- \rightarrow q \bar{q}~(\mu^+\mu^-)$  ($\sqrt{s}\approx 500\,$GeV), and  $10^6$ for e$^+$e$^- \rightarrow~$W$^+$W$^-$ ($\sqrt{s}\approx 500\,$GeV).
Hence, a statistical precision  at the per-mille level is expected.
The uncertainties of cross section measurements are not determined by statistics only but also by the precision of luminosity and energy measurement as well as polarization measurement,
\begin{equation}
\frac{\Delta \sigma}{\sigma} \propto \frac{1}{\sqrt{N}}\oplus \frac{\Delta {\cal{L}}}{{\cal{L}}}\oplus \frac{\Delta E}{E} \oplus \frac{\Delta P}{P}\,.
\end{equation}
Stability and measurement  of energy and luminosity  are aimed  below $10^{-4}$.
The error of the polarization measurement will be approximately 
$\Delta P_{e^+}/ P_{e^+} = \Delta P_{e^-}/ P_{e^-} = 0.25$\% (see~\cite{ref:jenny}).
The error of the effective  polarization for s-channel processes, $P_{eff}=(P_{e^+}-P_{e^-})/(1-P_{e^+}P_{e^-})$, is reduced considerably if both beams are polarized as can be shown simply by error propagation. 
 So, high energy measurements at the per-mille level are realistic but additional systematic effects would limit the precision measurements and hence the physics potential substantially. 

\section{Reversal of positron helicity}
What is the problem with the positron polarization? 

Let's consider for illustration the s--channel process e$^+$e$^- \rightarrow f \bar{f}$. If both beams are polarized four independent measurements can be performed:
\begin{eqnarray}
\sigma_{\pm \pm} & = & \frac{1}{4} \sigma_0 \left[ 1+P_{e^+}P_{e^-} +A_{LR}
\left(\pm P_{e^+} \pm P_{e^-} \right) \right] \label{eq:nonSM} \\
\sigma_{\mp \pm} & = & \frac{1}{4} \sigma_0 \left[ 1 -P_{e^+}P_{e^-} +A_{LR}
\left(\mp P_{e^+} \pm P_{e^-} \right) \right] \label{eq:SM}
\end{eqnarray}
Assuming 100\% polarization, $\pm P_{e^-} = \mp P_{e^+} = 1$, the cross sections (\ref{eq:nonSM}) are zero, and only  the Standard Model cross sections (\ref{eq:SM}) contribute. The Left-Right asymmetry, $A_{LR}$, can be determined from the measured left-right asymmetric cross sections with
\begin{equation}
A_{LR}= \frac{\sigma_{- +}- \sigma_{+ -}}{\sigma_{- +}+ \sigma_{+ -}} \cdot
       \frac{1-P_{e^+}P_{e^-}} {P_{e^+}-P_{e^-}}
=\frac{A_{LR}^{meas}}{P_{eff}}\label{eq:alr}
\end{equation} 
The effective polarization, $P_{eff}$, is larger than the individual polarizations $P_{e^+}$ and $P_{e^-}$, and due to error propagation,  $\Delta P_{eff}$ is smaller than the individual uncertainties $\Delta P_{e^+}$ and $\Delta P_{e^-}$. The measurement of the 
unpolarized cross section, $\sigma_0$,
\begin{equation}
\sigma_0 = \frac{1}{2} \cdot \frac{\sigma_{- +} +\sigma_{+ -}}{1-P_{e^+}P_{e^-}}\,,\label{eq:xs0}
\end{equation} 
demands a very precise knowledge of the beam polarizations and luminosity. 
The enhancement factor, $1-P_{e^+}P_{e^-}$, increases effectively the luminosity, e.g.\ by 24\% for $P_{e^+}=30$\% and $P_{e^-}=80$\%. 

To perform these measurements with the required precision, 
the luminosities delivered to the '+ $-$' and '$-$ +'  helicity states have to be equal, ${\cal{L}}_{+ -}={\cal{L}}_{- +}={\cal{L}}$, and equations (\ref{eq:alr}), (\ref{eq:xs0}) yield
\begin{equation}
\sigma_0 = \frac{1}{2} \cdot \frac{N_{- +} +N_{+ -}}{{\cal{L}} (1-P_{e^+}P_{e^-})}, ~~~~~A_{LR}= \frac{N_{- +}- N_{+ -}}{N_{- +}+ N_{+ -}} \cdot
       \frac{1}{P_{eff}}
\end{equation}
Same luminosities for each initial helicity combination
can be reached by frequent reversal of the helicit; so also time dependent fluctuations cancel. 
The reversal of the electron helicity is possible from train to train, $\pm P_{e^-}$.
For the undulator based positron source the reversal of the positron helicity is more complicated, since it depends on the orientation of helix windings in the undulator, i.e.\ only one polarization, $+ P_{e^+}$ or $- P_{e^+}$, can be produced.
 Opposite polarizations can be achieved 
by reversing the magnetic field in the spin rotators but this is only possible after a (few) run(s). As a consequence, one  luminosity is distributed to  electron--positron initial states with '$+ ~-$' and '$-~-$' helicity, another luminosity to '$+~+$' and '$-~+$'.
Although luminosity and energy should be stable at the $10^{-4}$ level,  equal luminosities for different runs are very hard to achieve.
Further, also the polarizations have to be the same for each run:  The
polarizations $P_{e^-}$ and $P_{e^+}$ occur in equations (\ref{eq:nonSM}) and (\ref{eq:SM}) linearly and bi-linearly. This complicates even tiny corrections substantially and requires also the knowledge of correlations. Long term variations in the machine and detector performance can add further sources for polarization dependent systematic uncertainties which can be traced only with fast helicity reversal. 

But also if one considers the ideal case of perfect reproducible and equal luminosities and polarizations in each run, a less frequent helicity reversal is a drawback: As mentioned, the effective luminosity of the interesting s--channel processes are enhanced by a factor
$1- P_{e^+}P_{e^-}$. This enhancement is fully lost if the positron helicity is less frequent flipped than the electron helicity because half of the running time will be spent for measurements of the 'inefficient' processes given in equation (\ref{eq:nonSM}).

Finally, the running strategy and distribution of luminosity to the helicity combinations '+ +', '$-$ $-$', '$-$ +' and '+ $-$' will depend on the physics requirements:
\begin{itemize}
\item  
If new physics beyond the Standard Model would enhance the cross sections (\ref{eq:nonSM}) these processes can be studied with running at  '+ +' and '$-$ $-$'.
\item
The analysis of all four independent measurements  (\ref{eq:nonSM}), (\ref{eq:SM}) allows a simultaneous measurement of $A_{LR}$,  $P_{e^+}$ and $P_{e^-}$~\cite{ref:blondel}. 
Especially in this case relative  fluctuations in polarization and luminosity  have to be determined and controlled.
A famous example is the GigaZ option with running the linear collider at the Z resonance~\cite{ref:gigaZ}. 
\end{itemize}
It would not be a good solution to destroy the positron polarization: 
The storage period of the beam in the damping ring is too short to delete the polarization~\cite{ref:larisa} and an additional facility would be needed. Also the zero polarization at the interaction point has to be confirmed by precise measurement.
  
A fast reversal of the positron helicity can be realized using fast kickers sending the positron beam to two spin rotator lines with opposite final spin orientation. Design proposals  for fast spin flip at energies of 5\,GeV or 400\,MeV are presented in references~\cite{ ref:flip5000,ref:flip400}.

\section{Conclusion}
It is an advantage to have a polarized positron beam 
from the beginning of  ILC operation. The undulator based positron 
source will provide $P_{e^+} > 30\%$, but this polarization can only be 
exploited for precision physics if a
flexible and fast helicity reversal for the positrons will be
available. Otherwise the slow helicity reversal would substantially
reduce or even cancel the advantage of having both beams polarized.   
After years of running the LHC it will be appropriate to operate a
linear collider that allows  flexible and comprehensive precision
studies of physics processes with  known initial states. 

\section*{Acknowledgments}
This work was supported in part by the Commission of the European Communities under the 6th Framework Programme ``Structuring the European Research Area'', contract number RIDS-011899.

\begin{footnotesize}

\end{footnotesize}
\end{document}